# BOSS: a tool for batch job monitoring and book-keeping

C. Grandi
*INFN Bologna, Viale Berti-Pichat 6/2, 40127 Bologna, ITALY*

BOSS (Batch Object Submission System) has been developed to provide real-time monitoring and bookkeeping of jobs submitted to a compute farm system. The information is persistently stored in a relational database (MySQL in the current version) for further processing. By means of user-supplied filters, BOSS extracts the specific job information to be monitored from the standard input, output and error of the job itself and stores it in the database in a structured form that allows easy and efficient access.
BOSS has been successfully used by all CMS Regional Centers for managing Monte Carlo data productions in 2002. Furthermore in fall 2002 it has been used in a prototype of the CMS production system deployed on the European DataGrid test bed demonstrating its ability to be used also in a grid environment.

## 1. INTRODUCTION

The management of a big Monte Carlo production or data analysis as well as the quality assurance of the results, require a careful monitoring and book-keeping of the batch jobs. This is normally achieved by parsing the *log file* of the job (i.e. the standard output and standard error of the job executable) with ad-hoc programs that can build summaries and/or check that the execution was as expected.

BOSS (Batch Object Submission System) has been developed to provide real-time monitoring and bookkeeping of jobs submitted to a compute farm system. The information is persistently stored in a relational database (MySQL in the current version) for further processing. In this way the information that was available in the *log file* in a free form is structured in a fixed-form that allows easy and efficient access. BOSS can monitor not only the typical information provided by the batch systems (e.g. executable name, time of submission and execution, return status, etc…), but also information specific to the job that is being executed (e.g. dataset that is being produced or analyzed, number of events done so far, number of events to be done, etc…).

BOSS extracts the specific job information to be monitored from the standard input, output and error streams of the job itself and stores it in the database. The fact that the information is extracted from the job rather than asserted by the user allows reliable bookkeeping. The user can provide BOSS with description of the parameters to be monitored and the way to access them by registering a *job type*. A *job type* is defined by:

- a *schema*, which describes the parameters to be monitored;
- a set of executables that contain the algorithms to determine the values of the parameters from the standard input, output and error of the job. The executables are the *pre-process*, *post-process* and *runtime-process* filters, which are executed before, after and during the job execution respectively.

Thus we can say that a user-job is of type *job type* if the *pre-process*, *post-process* and *runtime-process* filters can parse its standard input, output and error and determine the values of the parameters contained in the *schema*.

When a job is submitted through BOSS the request details, including its *job type*, are stored in the database. Then BOSS builds a wrapper around the job itself, the *jobExecutor*, which is submitted to the farm batch system together with all the information needed to run the user-job. When the execution starts a few processes are started on the execution host together with the real user executable. In particular the *dbUpdator* process that has the capability to submit queries to the relational database is used to update the job status in the database in real time. BOSS basic flow is shown in figure 1.

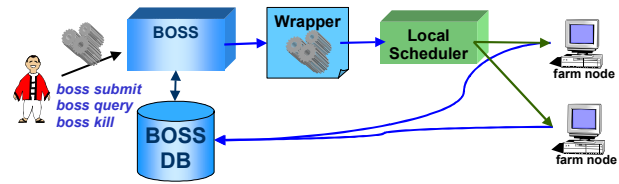

Figure 1: Basic flow of BOSS.

BOSS is not a batch system. It interfaces to a local scheduler (e.g. LSF, PBS, Condor, etc…) through a set of scripts provided by the farm administrator, using a predefined syntax. The local scheduler job identification string is kept in the database together with the BOSS job identification number so that it is always possible to reach the running job via the local batch system through BOSS commands. In the current version BOSS provides an interface to the local scheduler for the operations of job submission, deletion and query.

BOSS provides a few commands for browsing the database. They can be used inside user scripts to access in an easy way the information about the jobs. In this way the user can track the state of the running jobs as well as of the finished jobs, which allows identifying anomalies and building summaries.

BOSS has a command-line interface. Detailed description on its use may be found in [1] or at the BOSS web site [2].

## 2. ARCHITECTURE

The main components of BOSS are:
- The `boss` executable: the BOSS interface to the user
- The database: where BOSS stores job information

- The `jobExecutor` executable: the BOSS wrapper around the user-job
- The `dbUpdator` executable: the process that writes to the database as the job is running
- The local scheduler: the compute farm batch system

## 2.1. Interaction with the database

In the current release BOSS uses MySQL as underlying database system.

At installation time the user may modify the default database contact information that is written in a configuration file in the BOSS top directory. The file is read when compiling the BOSS executables. Since the executables are statically linked, the `boss` and the default `dbUpdator` executables are self-contained and connected to a single instance of a database. The `jobExecutor` executable is also statically linked but it doesn't depend on the database contact information.

The operations BOSS performs on the database are:
`CREATE TABLE`, `DROP TABLE`
`INSERT`, `DELETE`, `UPDATE`
`SELECT`, `SHOW FIELDS`
The data types used are:
`INT` (used for integers and time in UNIX format)
`VCHAR` (used for BOSS-defined strings)
`BLOB` (used for user-defined strings and for files, being them binaries or ASCII)

Any database system can be used in place of MySQL provided it supports these functionalities.

## 2.2. Interaction with the local batch system

From the architectural point of view it is possible to consider BOSS as the component that prepares the wrapper around the user-job letting the user submit the job to the preferred batch manager. Nevertheless submitting the job through the BOSS commands allows the complete tracking of its status because at submission time BOSS stores information in the database. Furthermore the job identifier assigned by BOSS to any job it manages (`BOSSJobId`) is put in one-to-one correspondence with the job identifier assigned to the job by the local batch system (`nativeJobId`) allowing further interaction with the job.

The interaction through BOSS with the local batch system is simple so that it is portable to a wide variety of implementations. Basically only three operations are considered: job submission, job cancellation and job status retrieval. Any other operation has to go directly through the batch system user interface. Support to a given batch system is provided via a set of executables that are registered in the form of plug-in to BOSS: `submit`, `query` and `kill`. The job submission plug-in only has to submit the BOSS wrapper (`jobExecutor`) with a few arguments, the most important being the `BOSSJobId` and has to return the job identifier assigned by the batch system to the job. The BOSS wrapper will then take care of executing the real job submitted by the user with the appropriate standard I/O streams and arguments.

Job cancellation requests are simply forwarded to the batch system.

The job status retrieval plug-in has to return a list of all the batch jobs known to it that are either running or queued, with a job status flag.

BOSS entirely relies on the batch system to execute the user-job reliably.

The plug-in executables are stored in the BOSS database in a dedicated table.

## 2.3. Job polymorphism

A "job" in BOSS architecture is characterized by a set of data variables that specify the job status and a set of methods to determine the status variables. Some of the data variables and the methods to determine them are common to any kind of job. These variables are typically those related to the batch system operation (submission time, executable name, standard I/O files, etc…) and are persistently stored in the BOSS database in a dedicated table that has the BOSS job identifier as primary key. Some other variables instead are typical of specific job types (e.g. number of simulated events produced by a given Monte Carlo job, non-standard I/O files used, etc…) and are determined by parsing the standard I/O streams of the job itself. In other words the user will deal with *concrete jobs* but BOSS will manage *abstract jobs*.

Even though BOSS is written in an Object Oriented language (C++), polymorphism is not implemented using inheritance since the definition of a new concrete job type should not imply recompilation of the BOSS executables. The way that is chosen is similar to that used for batch scheduler management:

Data variables are defined by means of a schema i.e. a comma separated list of `variable:type` pairs. Methods used to extract the variable values from the standard I/O streams are implemented as plug-in executables with defined interfaces that are invoked by BOSS (*filters*).

When a new job type is registered with BOSS, the schema and the plug-in executables are stored in the BOSS database in a dedicated table. Furthermore a new table is created in which the columns are the variables specified in its schema. In addition a column with the BOSS job identifier is added so that it is possible to link information stored in the specific job table to those in the standard job table.

The filters read the standard input, output and error streams of the user-job and write lists of *<variable=value>* pairs.

It is possible to associate to a single job more than one job type. If more job types are specified all the filter files of the different job types will parse the job standard I/O streams.

It is worth stressing that the only communication channels between the jobs and the BOSS system are the job I/O streams through the plug-in filters. In this view we can give this definition: *"a job is of a given type if the filters of that type are able to extract the value of the variables of that job-type from the job I/O streams"*.

When a job is submitted to BOSS, the user may optionally specify its *types*. BOSS retrieves from the database the filter files and stores them in an *archive file* together with the information needed to run the job (executable, standard I/O streams, arguments, etc...). The archive file must be made available to the executing job.

Figure 2 shows an example of the filtering procedure. The user-job writes to its standard output a string "`counter` *N*" every second, where *N* is the number of seconds since the beginning. The filter script looks for lines in the standard output matching this kind of string and writes out the string "`COUNTER=`*N*" where `COUNTER` is the name of the variable in the job schema (i.e. in the database table). The `jobExecutor` reads the output of the filter and writes to a journal file. Optionally the `jobExecutor` may start the `dbUpdator` process that updates the database.

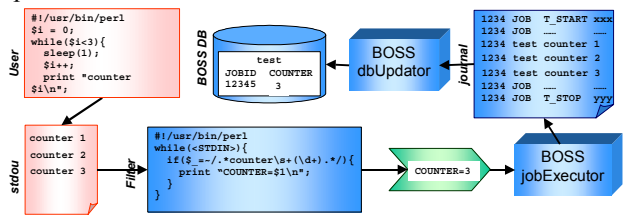

Figure 2: Simplified view of the BOSS filtering procedure

## 2.4. The job wrapper: `jobExecutor`

The executable that is started by the farm scheduler on a farm node is the BOSS wrapper (`jobExecutor`). The different steps performed by the `jobExecutor` are the following:

- If a *top working directory* is passed to the `jobExecutor`, a dedicated working directory is created under it.
- An empty *journal file* is created in the local working directory. It will contain all the updates to be done to the database.
- If real-time update of the database is requested the `dbUpdator` process is started. The `dbUpdator` is statically compiled with all the information needed to contact the database server hard-coded in it. The `dbUpdator` reads the journal file at regular intervals and updates the job information in the database.
- The details of the user-job (executable name, standard I/O stream files, etc...) are retrieved from the archive file that is found in the *base directory*.
- Pre-processing step: the user-specified standard input file is parsed by the `pre-process` filter files. The output of the filter is added by `jobExecutor` to the journal file.
- Run-time processing step: a number of temporary named pipes (FIFO's) and processes are created on the local farm node together with the real user executable. The standard output and error streams are intercepted by BOSS and processed using the job-type filters. The detailed schema of files and processes is shown in figure 3. The user-job is started using as standard input the file specified by the user. The standard output and error streams are sent to two FIFO's that are read by a "splitter" (the UNIX `tee` executable). One stream from the splitters is sent to the user-specified standard output and error files. Other streams (one for each job type) are created and sent to other FIFO's that are processed by the `runtime-process` filter files. The output of the filters is sent to another FIFO that is read by the `jobExecutor` itself that writes it to the journal file.
- Post-processing step: the standard output and error files, produced in the previous step are parsed by the `pre-process` filter files. The output of the filter is added by `jobExecutor` to the journal file.

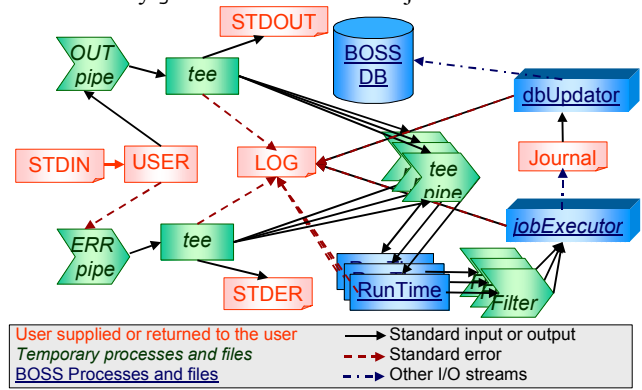

Figure 3: Schema of the processes and files created by `jobExecutor` at run-time

## 2.5. The user interface: `boss`

The `boss` executable is a command line interface to a set of commands. Each command is a C++ class that inherits from the `BossCommand` class that defines the interface. It is also possible to use the boss commands through their C++ API. Commands allow the user to:

- manage the schedulers registered to BOSS;
- manage the job types (schemas and filters) registered to BOSS;
- submit, terminate and get information about jobs;
- retrieve, update and remove information about jobs in the database.

## 2.6. Using ClassAd

ClassAd (Classified Advertisements) [3] is a syntax that allows specifying a mapping from *attribute names* to *expressions*. In the simplest cases, the expressions are simple constants (integer, floating point, or string). A ClassAd is thus a form of *property list*. In BOSS ClassAd may be used at submission time to specify the properties of a job in a file instead of specifying them as options on the `boss submit` command line. The advantage in using ClassAd's is that it is possible to add attributes (i.e.

name=value pairs) that are not standard BOSS submission options. These attributes are ignored by BOSS and passed as they are to the `submit` executable of the local batch system. Currently BOSS understands only the syntax of the Condor batch system and of the European DataGrid scheduler (JDL, see chapter 3). This means that a command file that is prepared following either Condor rules or of JDL can directly be submitted to BOSS. In principle it is possible to modify BOSS to understand any ClassAd based syntax.

## 3. USE IN GRID ENVIRONMENT

On the grid the submitting machine environment is completely disjoint from that of the executing machine. Since the MySQL server contact details are hard coded in the `dbUpdator` and that the `jobExecutor` and the `dbUpdator` itself are statically linked, it is possible to move them on any machine with the appropriate operating system and they will run. The only requirement in case real-time monitoring is required, is that the executing host has outbound connectivity and that the MySQL server accepts connections from all the grid nodes.

BOSS has been used with the scheduler provided by the (EDG) European DataGrid project [4]. The BOSS ClassAd parser has been modified to be able to understand the JDL (Job Description Language) syntax that is used by EDG to describe a job to be submitted. The BOSS archive file and all the needed files (i.e. the user specified executable file, the standard input file, the `jobExecutor`, the `dbUpdator`) are shipped to the execution host via the EDG *input sandbox* mechanism, and the BOSS journal file is shipped back in the *output sandbox* together with the specified standard output error and log files. All the communication from the worker node is through MySQL. This is understood to be a weakness of the system. To make the system more robust against network or server failures, a `dbUpdator` based on R-GMA (Relational Grid Monitoring Architecture) [5] is being developed. Figure 4 shows the process of remote update using R-GMA.

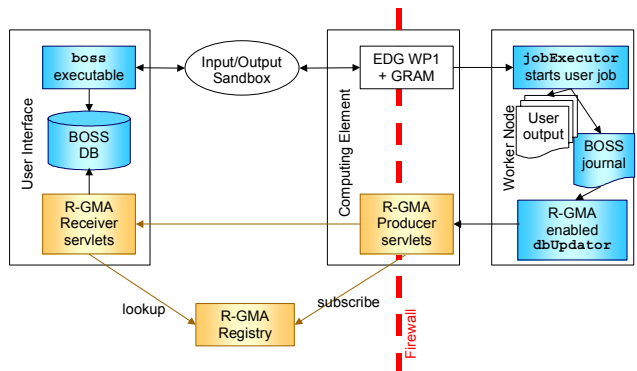

Figure 4: Use of R-GMA as a transport layer for BOSS

When the `dbUpdator` process starts on the Worker Node, it registers itself with the R-GMA *registry* as a *producer*. The operation is mediated by the producer *servlet* on the gatekeeper of the grid Computing Element. It is thus possible to run jobs also on Worker Nodes without outbound connectivity. Close to the BOSS database a *consumer* process is started. It finds out which jobs are running (or did run) from the registry and then it retrieves the information directly from the Computing Elements and stores them in the database. Since the job information is cached on the producer servlet on the Computing Element for an appropriate amount of time, it is possible to retrieve it also if the job is already finished and thus after a network or MySQL server interruption.

## 4. CURRENT USE OF BOSS

Previous versions of BOSS have been used as a job-monitoring tool by the CMS experiment during *Spring 2002* and *Summer 2002* data productions [6]. Every CMS Regional Center had a local MySQL server hosting the BOSS database. About 20 regional centers participated with about one thousand CPU's in total. Production ran continuously for about 4 months and over 100,000 jobs where executed using BOSS. At the time of writing, CMS is still producing simulated data and BOSS is also used for job monitoring.

A pre-release of BOSS version 3.3 has recently been used for monitoring CMS production jobs being submitted to the European DataGrid test-bed during the CMS-DataGrid stress test [7].